\documentclass[pra,twocolumn,showpacs,superscriptaddress,floatfix]{revtex4-1}  % for review and submission

\usepackage{graphicx}% Include figure files
\usepackage{dcolumn}% Align table columns on decimal point
\usepackage{bm}% bold math
\usepackage[mathlines]{lineno}% Enable numbering of text and display math
\usepackage{mathtools}
\usepackage{braket}
\usepackage{amsmath}
\usepackage{newtxtext}
\usepackage{verbatim}
\usepackage{xcolor}

\usepackage{accents}
\newcommand{\dbtilde}[1]{\accentset{\approx}{#1}}

\renewcommand{\vec}[1]{{\bm{\mathrm{#1}}}}
\newcommand{\vhat}[1]{{\hat{\bm{\mathrm{#1}}}}}

\newcommand{\RN}[1]{%
	\textup{\uppercase\expandafter{\romannumeral#1}}%
}

\let\Im\undefined
\DeclareMathOperator{\Tr}{Tr}

\DeclareMathOperator{\Im}{Im}

\DeclarePairedDelimiter \abs{\lvert}{\rvert}

\begin{document}

\preprint{APS/123-QED}

\title{Role of orbital hybridization in anisotropic magnetoresistance}

\author{Hye-Won Ko}
\affiliation{KU-KIST Graduate School of Converging Science and Technology, Korea University, Seoul 02841, Korea}

\author{Hyeon-Jong Park}
\affiliation{KU-KIST Graduate School of Converging Science and Technology, Korea University, Seoul 02841, Korea}

\author{Gyungchoon Go}
\affiliation{Department of Materials Science and Engineering, Korea University, Seoul 02841, Korea}

\author{Jung Hyun Oh}
\affiliation{Department of Materials Science and Engineering, Korea University, Seoul 02841, Korea}

\author{Kyoung-Whan Kim}
\email{kwk@kist.re.kr}
\affiliation{Center for Spintronics, Korea Institute of Science and Technology, Seoul 02972, Korea}

\author{Kyung-Jin Lee}
\email{kj_lee@korea.ac.kr}
\affiliation{KU-KIST Graduate School of Converging Science and Technology, Korea University, Seoul 02841, Korea}
\affiliation{Department of Materials Science and Engineering, Korea University, Seoul 02841, Korea}

\date{\today}% It is always \today, today, but any date may be explicitly specified

\begin{abstract}
We theoretically and numerically show that longitudinal orbital currents in ferromagnets depend on the magnetization direction, which contribute to the anisotropic magnetoresistance (AMR). This orbital contribution to AMR arises from the momentum-dependent orbital splitting, which is generally present in multi-orbital systems through the orbital anisotropy and the orbital hybridization. We highlight the latter orbital hybridization as an unrecognized origin of AMR and also as a common origin of AMR and orbital Hall effect.
%\begin{description}
% Secondary publications and information retrieval purposes.
% You may use the \texttt{description} environment to structure your abstract;
% use the optional argument of the \verb+\item+ command to give the category of each item.
%\end{description}
\end{abstract}


\maketitle

% \section{\label{sec:level1}First-level heading}
% \subsection{\label{sec:level2}Second-level heading: Formatting}
% \subsubsection{Wide text (A level-3 head)}

%%%%%%%%%%%%%%%%%%%%%%%%%%%%%%%%%%%%%%%%%%%%%%%%%%%%%%%%%%%%%%%%%%%%%%%

\section{\label{sec:Intro}Introduction}
The anisotropic magnetoresistance (AMR), the dependence of electrical conductivity on the orientation of magnetization with respect to the electric current direction, was first reported by Thomson~\cite{Thomson1857}. A concurrent action of magnetization and spin-orbit coupling (SOC) accounts for occurrence of the anisotropic conduction. Smit~\cite{Smit1951} suggested that an admixture of parallel and antiparallel \textit{d} states due to SOC results in an unequal distribution of electrons in \textit{d} orbitals, in which the inequality is determined by the direction of magnetization. This suggestion was further reinforced~\cite{Campbell1969} by a two-current model with \textit{s-s} and \textit{s-d} transitions regarding a perturbation owing to SOC, $(L_+S_-+L_-S_+)/2$ where $L_\pm(S_\pm)=L_x(S_x)\pm i L_y(S_x)$ corresponds to orbital (spin) angular momentum operator. Another mechanism, proposed by Berger~\cite{Berger1964}, considered the effect of $L_zS_z$ which causes an anisotropic shape of 3\textit{d} atomic wave functions. Subsequent studies have led to profound understanding on AMR (for a detailed review see Ref.~\cite{Mcguire1975}). One of these works~\cite{Velev2005} introduced orbital degrees of freedom which correlate with the magnetization through a concerted action between the exchange coupling and SOC. This correlation, given as $\pm\frac{1}{2}\lambda L_\vec{m}$ where $\lambda$ is the SOC strength and $L_\vec{m}$ denotes the orbital angular momentum operator projected on the magnetization direction $\vec{m}$, splits the orbital energies according to the orientation of magnetization. Considering ballistic transport in ferromagnetic nanowires, they showed that the anisotropic conductance originated from magnetization-dependent electronic structure at the Fermi energy.

Since the orbital degrees of freedom are coupled to the spin degrees of freedom through the SOC, the impact of orbitals in SOC-related phenomena is an important task to investigate. For instance, the surface states of topological insulators have a chiral spin texture in momentum space~\cite{Hsieh2009a,Hsieh2009b} which fosters interesting consequences such as Edelstein effect~\cite{Edelstein1990}, prohibition of backscattering~\cite{Roushan2009}, and so forth. Experimental~\cite{Xie2014,Park2012} and theoretical~\cite{Zhang2013,Park2012} efforts have verified the existence of analogous momentum-space chiral orbital texture that is coupled to the configuration of spins in momentum space through the SOC. Likewise, the spin-momentum locking in Rashba-type band structure is also derived from the orbital Rashba effect~\cite{Park2011,Go2017} which is prior to the spin Rashba effect. The importance of orbital degrees of freedom for spin-transfer torques was also highlighed~\cite{Paul}. Considering the close connection between orbital and spin degrees of freedom, an exploration on orbital-related phenomena is significant to elucidate the underlying mechanism of corresponding spin-related-phenomena in spin-orbit-coupled systems.

Orbital textures in momentum space exist even in topologically trivial and centrosymmetric systems and thus are quite generic~\cite{Tanaka2008,Kontani2009,Go2018}. This owes to the orbital hybridization, which is an overlap between orbitals with distinct angular quantum numbers in neighboring atomic sites and is a general property of multi-orbital systems. An important outcome of the orbital hybridization is the orbital Hall effect which refers to the transverse orbital current induced by electric field. The orbital Hall effect has been remarked as the origin of intrinsic spin Hall effect. Kontani \textit{et al}.~\cite{Kontani2009} demonstrated that the orbital Aharonov-Bohm phase arose from \textit{sd} hybridization yields the giant orbital Hall effect, resulting in spin and anomalous Hall effects. In addition, Go \textit{et al}.~\cite{Go2018} systematically investigated the dependence of both intrinsic spin Hall and orbital Hall conductivities on orbital hybridization strength and emphasized the significance of the orbital hybridization in spin-orbit coupled transport.

In this paper, we theoretically analyze the AMR in terms of orbital degrees of freedom. We show that the longitudinal orbital conductivity also depends on the magnetization direction, which we call {\it orbital anisotropic magnetoresistance} (OAMR), as the charge conductivity does. The magnetization-dependent conductivities come from the momentum-dependent orbital splitting which is achieved by \textit{orbital anisotropy} or \textit{orbital hybridization}. The former was alluded as the symmetrical characteristics of each orbitals~\cite{Velev2005} while the latter is the newly found origin of AMR in this work. %We below mention how each of them is quantified by tight-binding parameters and we call these two factors orbital factors throughout this paper.

This paper is organized as follows. In Sec.~\ref{sec: OS}, we first present a simple model demonstrating how the oribtal hybridization alone induces the orbital splitting and causes the anisotropic conduction. An analytic derivation of magnetization-dependent conductivity is also shown by perturbation theory in Sec.~\ref{sec: Kubo}. Then in Sec.~\ref{sec: Num}, we numerically compute the charge, spin, and orbital conductivities based on Green's function formalism~\cite{Ghosh2018} with the magnetization parallel and perpendicular to the current direction where the two orbital factors, the orbital anisotropy and orbital hybridization, are treated independently. We also inspect the SOC strength dependence of OAMR and AMR to examine the connection between them. This work suggests that AMR is closely related to OAMR where the role of orbital hybridization is signified as the underlying mechanism. It is noted that we focus on the mechanism developed by the anisotropic band structures but ignore the process related to the anisotropic scattering, which is manifested as the magnetization-dependent relaxation time.

\section{\label{sec: OS}Orbital splitting due to orbital hybridization}

\begin{figure}[t]
\begin{center}
\includegraphics[scale=0.3]{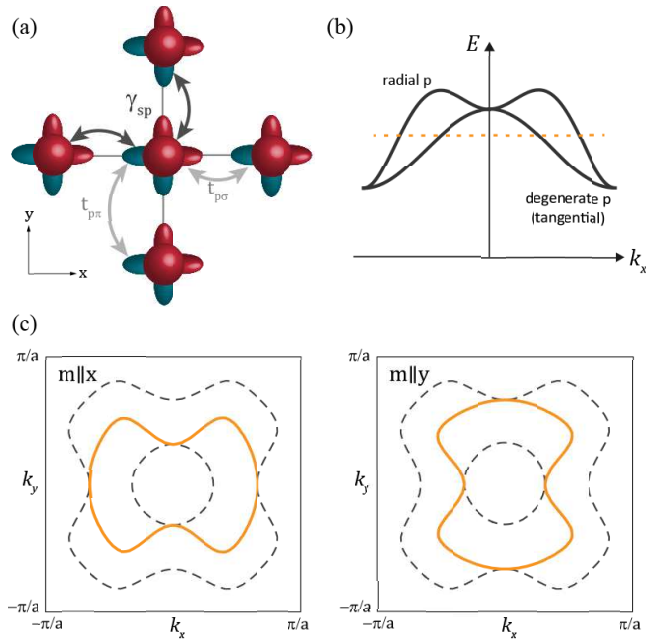}
\caption{
(a) An illustration of tight-binding model in a cubic lattice with $s$, $p_x$, $p_y$, and $p_z$ orbitals. The hopping integrals between \textit{p} orbitals (lighter arrows) are $t_{p\sigma}$ and $t_{p\pi}$ according to relevant bonding type and the hopping between \textit{s} and \textit{p} orbitals (darker arrows) is represented as $\gamma_{sp}$. Note that the hopping between neighboring $s$ orbitals is omitted. (b) Schematic band structure showing lifted degeneracy of \textit{p}-orbitals originating from the inclusion of orbital hybridization [Eq.~(\ref{Eq2})]. (c) Fermi surface (${k_z}=0$) at an energy crossing $p$ orbital energies [dotted line in (b)]. Dashed contours are the energies without effective SOC, $\lambda\vec{L}\cdot\vec{m}$. When the magnetization direction is determined, each orbitals experience orbital-dependent spin-orbit splitting. Solid contours are the resultant Fermi surface when (left panel) $\vec{m}\parallel\vhat{x}$ and (right panel) $\vec{m}\parallel\vhat{y}$.  
}
\label{fig:FIG1}
\end{center}
\end{figure}

We start with a simple tight-binding model, consisting of $s$ and three $p$ orbitals ($p_x$, $p_y$, $p_z$) in a cubic lattice [Fig.~\ref{fig:FIG1}(a)]. The interatomic hopping integrals can be classified by the equivalence of orbital types involved in hopping between neighboring atomic sites. One is the hopping between identical types of orbital such as {\textit{s}-to-\textit{s}, \textit{p}-to-\textit{p}} [lighter arrows in Fig.~\ref{fig:FIG1}(a)] and the other is the hopping between different types of orbital, \textit{s}-to-\textit{p} and vice versa [darker arrows in Fig.~\ref{fig:FIG1}(a)]. Except for isotropic $s$ orbital, the strength of the former hopping strongly depends on its hopping direction (e.g., across $\sigma$ bond or $\pi$ bond), which results in anisotropic splitting of orbital energies. We call this former pathway providing an anisotropic degree to the system as the orbital anisotropy and describe it with the difference between $t_{p\sigma}$ and $t_{p\pi}$. The latter hopping between different orbitals is mediated by the hybridization between \textit{s} and \textit{p} orbitals and gives another degree of anisotropy (shown below). We specify this latter route as the orbital hybridization, and describe it with a hopping parameter between $s$ and $p$, $\gamma_{sp}$.

Let us first consider the simplest case where none of both orbital factors, exchange interaction, and SOC are present. In this case, the orbital part of the Hamiltonian is simplified as,
\begin{align}
{\cal{H}}_\mathrm{L} = \sum_{\sigma=\uparrow,\downarrow} \Big[&E_s(\vec{k})\ket{\vec{k}s^\sigma}\bra{\vec{k}s^\sigma} \nonumber \\
&+\sum_{i=x,y,z}E_{p_i}(\vec{k})\ket{\vec{k}p_i^\sigma}\bra{\vec{k}p_i^\sigma}\Big], \label{Eq1}
\end{align}
where $E_{s(p_i)}(\vec{k})$ is the energy of $s(p_i)$ orbitals ($i=x,y,z$) with a given Bloch momentum $\vec{k}$ and $\ket{\vec{k}\alpha^\beta}$ is the electronic state with the Bloch momentum $\vec{k}$, orbital state $\alpha$, and spin $\beta$. Here $\alpha=s,\ p_x,\ p_y,\ p_z$ and $\beta=\ \uparrow,\downarrow$. Note that $\ket{\vec{k}\alpha^\beta}\equiv\ket{\psi_{\alpha\beta}^\vec{k}}$ in Appendix~\ref{appendix: TB Ham}. Because none of orbital factors, exchange coupling, and SOC are present, the energies of $p$ orbitals are degenerate (i.e., $E_{p_x}=E_{p_y}=E_{p_z}$). Without loss of generality we can adopt spherical coordinates (see Appendix~\ref{appendix: Spher Coord}) and rewrite the second term in Eq.~(\ref{Eq1}) as the summation of the radial \textit{p} orbital contribution, $E_{pr}(\vec{k})\ket{\vec{k}p_\vec{k}^\sigma}\bra{\vec{k}p_\vec{k}^\sigma}$, and the tangential \textit{p} orbital contributions, $E_{pt}(\vec{k})(\ket{\vec{k}p_{\theta_\vec{k}}^\sigma}\bra{\vec{k}p_{\theta_\vec{k}}^\sigma}+\ket{\vec{k}p_{\phi_\vec{k}}^\sigma}\bra{\vec{k}p_{\phi_\vec{k}}^\sigma})$. Here, $p_\vec{k}$ is the radial \textit{p} orbital whose lobe is along $\vec{k}$ and $p_{\theta_\vec{k}/\phi_\vec{k}}$ is the tangential \textit{p} orbital whose lobe is perpendicular to $\vec{k}$. $E_{pr}(\vec{k})$ and $E_{pt}(\vec{k})$ are the energies of radial and tangential \textit{p} orbitals, respectively. Note that the degeneracy of \textit{p} orbitals is still maintained (i.e., $E_{pr}=E_{pt}$).

Consideration of the orbital factors breaks the degeneracy. For the orbital anisotropy, the hopping integral for the radial orbital is $t_{p\sigma}$ and that for the tangential orbital is $t_{p\pi}$, thus $E_{pr}$ and $E_{pt}$ become different. For the orbital hybridization, the hybridization between $s$ orbital and the radial orbital breaks the degeneracy between $E_{pr}$ and $E_{pt}$. Since the effect of the former on AMR was studied in Ref.~\cite{Velev2005}, we discard it in this section, in order to demonstrate that the orbital hybridization alone can result in the orbital splitting. The orbital anisotropy is restored in Sec.~\ref{sec: LRT}.

Now we turn on the orbital hybridization. The orbital hybridization energy $\gamma_{sp}(\vec{k})$ gives the \textit{sp} hybridization Hamiltonian,
\begin{align}
{\cal{H}}_{\mathrm{OH}}= \sum_{\sigma=\uparrow,\downarrow} i\gamma_{sp}(\vec{k})(\ket{\vec{k}s^\sigma}\bra{\vec{k}p_\vec{k}^\sigma}-\ket{\vec{k}p_\vec{k}^\sigma}\bra{\vec{k}s^\sigma}), \label{Eq2}
\end{align} 
where a finite overlap with $s$ orbital is obtained only for radial $p$ orbital. The $s$-to-$p$ hopping is annihilated for the tangential $p$ orbitals due to opposite contributions of each lobes [red and green lobes in Fig.~\ref{fig:FIG1}(a)] which represent opposite orbital phases. Restoring the orbital hybridization to the $sp$ system, i.e., ${\cal{H}}_\mathrm{L}+{\cal{H}}_\mathrm{OH}$, the degeneracy of $p$ orbitals is lifted according to its character of orbital wavefunction [upper (radial) and lower (tangential) energies in Fig.\ref{fig:FIG1}(b)]. The resulting energies of $s$ and $p$ states are,
\begin{align}
\tilde{E}_s(\vec{k}) =& \ \frac{E_s(\vec{k})+E_{pr}(\vec{k})}{2}-\epsilon(\vec{k}), \nonumber \\
\tilde{E}_{pr}(\vec{k}) =& \ \frac{E_s(\vec{k})+E_{pr}(\vec{k})}{2}+\epsilon(\vec{k}), \nonumber \\
\tilde{E}_{pt}(\vec{k}) =& \ E_{pt}(\vec{k}), \label{Eq3}
\end{align}
where $\tilde{E}_{\alpha'}(\vec{k})$ is an eigenvalue of $\cal{H}_\mathrm{L}+\cal{H}_\mathrm{OH}$ ($\alpha'=s, \ pr, \ pt$) and $\epsilon(\vec{k})=\sqrt{[E_s(\vec{k})-E_{pr}(\vec{k})]^2/4+\gamma_{sp}^2(\vec{k})}$. 

Then we bring back the exchange interaction and SOC. In the strong exchange limit of ferromagnet, a concerted action of the exchange interaction and SOC produces an effective correlation between orbital and magnetization~\cite{Velev2005}, $\cal{H}_\mathrm{SO}\sim\lambda\vec{L}\cdot\vec{m}$ [Eq.~(\ref{BEq4})]. Combining this effective SOC to the previous system gives rise to the magnetization-dependent energy splitting if the orbital splitting is present beforehand. For example when the magnetization is along the momentum direction (i.e., $\vec{m}\parallel\vec{k}$), the $p$ orbital energies of Hamiltonian $\cal{H}_\mathrm{L}+\cal{H}_\mathrm{OH}+\cal{H}_\mathrm{SO}$ are, 
\begin{align}
\dbtilde{E}_{pr}^{\parallel}(\vec{k}) =& \ \tilde{E}_{pr}(\vec{k}), \nonumber \\
\dbtilde{E}^{\parallel,\pm}_{pt}(\vec{k}) =& \ \tilde{E}_{pt}(\vec{k})\pm\lambda. \label{Eq4}
\end{align}
Here, $\dbtilde{E}_{\alpha'}^\parallel(\vec{k})$ is the eigenenergy of $\cal{H}_\mathrm{L}+\cal{H}_\mathrm{OH}+\cal{H}_\mathrm{SO}$ in which the subscript denotes the orbital character and the superscript $\parallel$ designates the relative orientation between the momentum $\vec{k}$ and the magnetization $\vec{m}$. An additional superscript of tangential $p$ orbitals corresponds to the sign of spin-orbit splitting, $\pm\lambda$. Note that we omit the energy of $s$ orbital as the $s$ orbital is the state for zero angular momentum. In contrast, if the magnetization is perpendicular to the momentum (i.e., $\vec{m}\cdot\vec{k}=0$), we obtain the $p$ orbital energies as, 
\begin{align}
\dbtilde{E}_{pr}^{\perp}(\vec{k}) &= \ \frac{\tilde{E}_{pr}(\vec{k})+\tilde{E}_{pt}(\vec{k})}{2}+\zeta(\vec{k}), \nonumber \\
\dbtilde{E}_{pt}^{\perp}(\vec{k}) &= 
	\begin{cases}
	\displaystyle\frac{\tilde{E}_{pr}(\vec{k})+\tilde{E}_{pt}(\vec{k})}{2}-\zeta(\vec{k}), \\
	\\
	\tilde{E}_{pt}(\vec{k}),
	\end{cases}
\label{Eq5}
\end{align}
where $\zeta(\vec{k})=\sqrt{[\tilde{E}_{pr}(\vec{k})-\tilde{E}_{pt}(\vec{k})]^2/4+\lambda^2}$ and a superscript $\perp$ denotes that $\vec{m}\perp\vec{k}$. Note that the upper (lower) case of tangential $p$ orbitals corresponds to the orbital state which is constructed with the eigenstates perpendicular (parallel) to the magnetization direction, e.g., $\ket{\vec{k}p_{\theta_\vec{k}}} \left(\ket{\vec{k}p_{\phi_\vec{k}}}\right)$ when $\vec{m}\parallel\vhat{\phi}_\vec{k}$. The discrepancy between Eq.~(\ref{Eq4}) and Eq.~(\ref{Eq5}) stems from the indirect coupling between orbital and magnetization, $\vec{L}\cdot\vec{m}$, through SOC.

The variation of magnetization direction modifies the form of effective SOC which affect the orbital energies in an orbital-dependent manner. However if the orbital splitting is absent (i.e., all orbitals are degenerate), the spin-orbit splitting results in identical orbital energies regardless of the magnetization direction, $E_{pr}\pm\lambda$ and $E_{pr}$. For this reason, the presence of orbital splitting is necessary for magnetzation-dependent and anisotropic band structures. Figure \ref{fig:FIG1}(c) shows the alteration of Fermi surface at $k_z=0$ for varied $\vec{m}$ direction, $\vec{m}\parallel\vhat{x}$ and $\vec{m}\parallel\vhat{y}$. Before the inclusion of $\cal{H}_\mathrm{SO}$ [dashed curves in Fig.~\ref{fig:FIG1}(c)], the inner and outer Fermi surfaces are developed due to the orbital hybridization, $\cal{H}_\mathrm{OH}$. Based on this orbital splitting, the effective SOC modifies the Fermi surfaces [solid curves in Fig.~\ref{fig:FIG1}(c)] which display magnetization-dependent band structures as Eqs.~(\ref{Eq4}) and (\ref{Eq5}). When an electric field is applied in the $x$ direction, the Fermi surfaces are shifted along the same direction. In consequence of the anisotropic band structures, an electric transport will also show the magnetization-dependent behavior, thus AMR.

As a side remark, the above simple argument works whenever there is an orbital splitting. Although we focus on the orbital hybridization contribution in this paper, the orbital anisotropy, for which $E_{pr}$ and $E_{pt}$ are different, also makes $\tilde{E}_{pr}$ and $E_{pt}$ different as studied in Ref.~\cite{Velev2005}.

\section{\label{sec: LRT}Linear Response Theory}
\subsection{\label{sec: Kubo}Analytical calculation}
In this section, we analytically show that the orbital hybridization gives rise to the AMR. The AMR is given by the magnetization-dependent part of the electrical conductivity. Basically, the problem is $8\times8$ since we have one \textit{s} orbital and three \textit{p} orbitals for each of spins. We assume that the exchange interaction is strong enough for spins to be aligned with the magnetization and thus reduce the problem to $4\times4$. %We also discard the orbital anisotropy for simplicity, since its effect has been already studied theoretically~\cite{Velev2005}.
First, we express our Hamiltonians in a spherical coordinate basis. To utilize the expression in Appendix~\ref{appendix: Spher Coord}, we choose our basis as $(\ket{\vec{k}s^\sigma},\ket{\vec{k}p_\vec{k}^\sigma},\ket{\vec{k}p_-^\sigma},\ket{\vec{k}p_+^\sigma})$, where  $\ket{\vec{k}p_\pm^\sigma}=(\ket{\vec{k}p_{\theta_\vec{k}}^\sigma}\pm i\ket{\vec{k}p_{\phi_\vec{k}}^\sigma})/\sqrt{2}$. In this basis,
\begin{equation}
\mathcal{H}_{\rm L}+\mathcal{H}_{\rm OH}=\begin{pmatrix}
E_s& -i\gamma_{sp} & 0 &0  \\
i\gamma_{sp} & E_{pr} & 0 & 0 \\
0 & 0 & E_{pt} & 0 \\
0 & 0 & 0 & E_{pt}
\end{pmatrix}.\label{Eq:HL+HOH}
\end{equation}
Hereafter we omit the $\vec{k}$ dependencies in the energies for concise presentation. Note that diagonalizing the upper $2\times2$ block gives the $sp$ hybridized states in Eq.~(\ref{Eq3}). Since the tangential \textit{p} orbitals are degenerate, the choice of the linear combination of the tangential \textit{p} orbitals, i.e. $(\ket{\vec{k}p_{\theta_\vec{k}}^\sigma}\pm i\ket{\vec{k}p_{\phi_\vec{k}}^\sigma})/\sqrt{2}$, does not affect the matrix representation of $\mathcal{H}_{\rm L}+\mathcal{H}_{\rm OH}$. However, it affects the SOC Hamiltonian significantly [Compare Eqs.~(\ref{BEq4}) and (\ref{BEq5})].

The SOC Hamiltonian in this basis is given by Eq.~(\ref{BEq5}).
\begin{equation}
\cal{H}_\mathrm{SO}=-\lambda\begin{pmatrix}
0 & 0 & 0 & 0 \\
0 & 0 & m_{-,\vec{k}} & im_{+,\vec{k}} \\
0 & m_{+,\vec{k}} & -m_\vec{k} & 0 \\
0 & -im_{-,\vec{k}} & 0 & m_\vec{k} \\
\end{pmatrix},\label{Eq:HSO}
\end{equation}
where $m_\vec{k}=\vec{m}\cdot\vhat{k}$, $m_{\pm,\vec{k}}=\vec{m}\cdot(\vhat{\theta}_\vec{k}\pm i\vhat{\phi}_\vec{k})/\sqrt{2}$, and the explicit expressions of the spherical coordinate vectors are presented in Eq.~(\ref{BEq1}). From the above expression, it is evident that the role of SOC is twofold. First, the radial component of magnetization ($m_\vec{k}$) breaks the degeneracy between the tangential orbitals [corresponding to Eq.~(\ref{Eq4})]. Second, the tangential components of magnetization $m_{\pm,\vec{k}}$ in the off-diagonal components of $\mathcal{H}_{\rm SO}$ mix the $sp$ hybridized states in the upper $2\times2$ block and the tangential orbitals in the lower $2\times2$ block [corresponding to Eq.~(\ref{Eq5})].

Now we treat SOC perturbatively and calculate its second order contribution to AMR. Our perturbation theory consists of two steps. At the first step, we block-diagonalize the $4\times4$ Hamiltonian to two $2\times 2$ blocks, and at the second step, we diagonalize each of the $2\times2$ blocks. For the first step, we decompose the Hamiltonian as $\mathcal{H}_{\rm tot}=\mathcal{H}_0+\mathcal{H}_1$, where
\begin{align}
\mathcal{H}_0&=\begin{pmatrix}
E_s& -i\gamma_{sp} & 0 &0  \\
i\gamma_{sp} & E_{pr} & 0 & 0 \\
0 & 0 & E_{pt}+\lambda m_\vec{k} & 0 \\
0 & 0 & 0 & E_{pt}-\lambda m_\vec{k}
\end{pmatrix}\nonumber\\
\mathcal{H}_1&=-\lambda\begin{pmatrix}
0 & 0 & 0 & 0 \\
0 & 0 & m_{-,\vec{k}} & im_{+,\vec{k}} \\
0 & m_{+,\vec{k}} & 0 & 0 \\
0 & -im_{-,\vec{k}} & 0 & 0 \\
\end{pmatrix}.
\end{align}
Here, $\mathcal{H}_0$ is the block-diagonal part and $\mathcal{H}_1$ is the block-off-diagonal part that is treated perturbatively. To block-diagonalize $\mathcal{H}_{\rm tot}$, we follow the Schrieffer–Wolff procedure~\cite{SWolff} by introducing a unitary transformation $\mathcal{H}_{\rm tot}'=e^\mathcal{S}\mathcal{H}_{\rm tot}e^{-\mathcal{S}}$ by choosing
\begin{align}
\mathcal{S}&=\begin{pmatrix}
0 & S_{2\times2} \\ 
-S_{2\times2}^\dagger & 0
\end{pmatrix} ,\nonumber\\
S_{2\times2}&=-\lambda f\begin{pmatrix}
-i\gamma_{sp}\\E_{pt}-E_s
\end{pmatrix}\begin{pmatrix}
m_{-,\vec{k}}&im_{+,\vec{k}}
\end{pmatrix}\nonumber\\
f&=[(E_{pr}-E_{pt})(E_{pt}-E_s)+\gamma_{sp}^2]^{-1}.\label{Eq:S}
\end{align}
Note that $\mathcal{S}$ satisfies $[\mathcal{H}_0, \mathcal{S}]=\mathcal{H}_1+\mathcal{O}(\lambda^2)$. Then, the Baker-Hausdorff Lemma gives the transformed Hamiltonian being
\begin{align}
\mathcal{H}_{\rm tot}'&=\begin{pmatrix}
H_{sp} & 0 \\ 
0 & H_{pt}
\end{pmatrix} +\mathcal{O}(\lambda^3),\nonumber\\
H_{sp}&=\begin{pmatrix}
E_s & -i\gamma_{sp} \\ 
i\gamma_{sp} & E_{pr}
\end{pmatrix}+\lambda^2(1-m_\vec{k}^2)f\begin{pmatrix}
0 & -\frac{i\gamma_{sp}}{2} \\ 
\frac{i\gamma_{sp}}{2} & E_{pt}-E_s
\end{pmatrix}\nonumber\\
H_{pt}&=E_{pt}+\lambda\begin{pmatrix}
m_\vec{k} & 0 \\ 
0 & -m_\vec{k}
\end{pmatrix}\nonumber\\
&\quad-\frac{\lambda^2f}{2}(E_{pt}-E_s)\begin{pmatrix}
1-m_\vec{k}^2 & 2im_{+,\vec{k}}^2 \\ 
-2im_{-,\vec{k}}^2 & 1-m_\vec{k}^2
\end{pmatrix}.\label{H_tot2}
\end{align}
Here we used $2m_{+,\vec{k}}m_{-,\vec{k}}=1-m_\vec{k}^2$. Note that $\mathcal{H}_{\rm tot}'$ is block diagonal, so one can diagonalize each of the $2\times2$ blocks ($H_{sp}$ and $H_{pt}$) to analytically obtain the perturbed energy eigenvalues $E_n(\vec{k})$ and the velocity (along $x$) $v_{n,x}(\vec{k})=(1/\hbar)\partial k_xE_n(\vec{k})$. Here $n=0,1,2,3$ is the band index for $s$ and $p$ states. Although we do not explicitly express $E_n(\vec{k})$ here, the second order correction is in the form of
\begin{equation}
\Delta E_n(\vec{k})=m_\vec{k}^2E_{n,2}(\vec{k}).\label{Eq:En}
\end{equation}
This magnetization-dependent correction is the key result of our perturbation theory.

To calculate the electrical conductivity, we ignore the interband contribution which is negligible when the level broadening is smaller than the energy differences between the bands. Then, the electrical conductivity is approximated as
\begin{equation}
\sigma=\frac{e^2\tau}{V}\sum_{\vec{k},n}\delta(E_F-E_n(\vec{k}))v_{n,x}(\vec{k})^2,\label{Eq:sigma}
\end{equation}
where $E_F$ is the Fermi level and $\hbar/\tau$ is the level broadening. The magnetization-dependent contribution in Eq.~(\ref{Eq:En}) clearly implies the existence of AMR in our model. To explicitly demonstrate that the AMR is generated by the orbital hybridization, {we discard the effect of orbital anisotropy ($E_{pr}=E_{pt}$) which has been already studied~\cite{Velev2005} and consider only the $s$ band is located at the Fermi level. Then, up to second order in $\gamma_{sp}$, the energy eigenvalue for $H_{sp}$ in Eq.~(\ref{H_tot2}) gives
\begin{align}
\dbtilde{E}_s(\vec{k})&\approx E_s-\frac{\gamma_{sp}^2}{E_{pr}-E_s}-\frac{(1-m_\vec{k}^2)\lambda^2\gamma_{sp}^2}{(E_{pr}-E_s)^3}\nonumber\\
&\equiv E_{0,0}(\vec{k})+m_\vec{k}^2E_{0,2}(\vec{k}),
\end{align}
where $E_{0,2}\propto\lambda^2\gamma_{sp}^2$. Assuming that $E_{0,0}(\vec{k})$ and $E_{0,2}(\vec{k})$ are spherically symmetric (thus dependent solely on $k$) and quadratic in $k$, Appendix~\ref{appendix: AMR} gives the electrical conductivity
\begin{equation}
\sigma=\frac{e^2\tau k_F}{3\hbar^2\pi^2}[E_F+2E_{0,2}(k_F) m_x^2],\label{Eq:AMR}
\end{equation}
where $k_F$ is the Fermi wave vector. The second term ($\propto\lambda^2\gamma_{sp}^2$) corresponds to the AMR induced by the orbital hybridization.

We remark that the function $f(\vec{k})$ in Eq.~(\ref{Eq:S}) diverges when both the orbital anisotropy and the orbital hybridization are zero. However, this does not mean that the SOC contribution diverges there. Since our perturbation theory treats $\lambda$ as the smallest factor, it is implicitly assumed that $\lambda\ll f^{-1/2}$, which is consistent with our numerical simulation parameters. If this assumption does not hold, one needs to start with the exact expression of $\mathcal{S}$ in Appendix~\ref{appendix: exact S}, expand $\mathcal{S}$ with respect to $\gamma_{sp}$ and $E_{pt}-E_{pr}$ first, and then discard $\mathcal{O}(\lambda^3)$. This procedure yields a different expression, but the main message of our theory is unaltered by the order of expansions.

\subsection{\label{sec: Num}Numerical calculation}

\begin{figure*}[t]
\begin{center}
\includegraphics[scale=0.45]{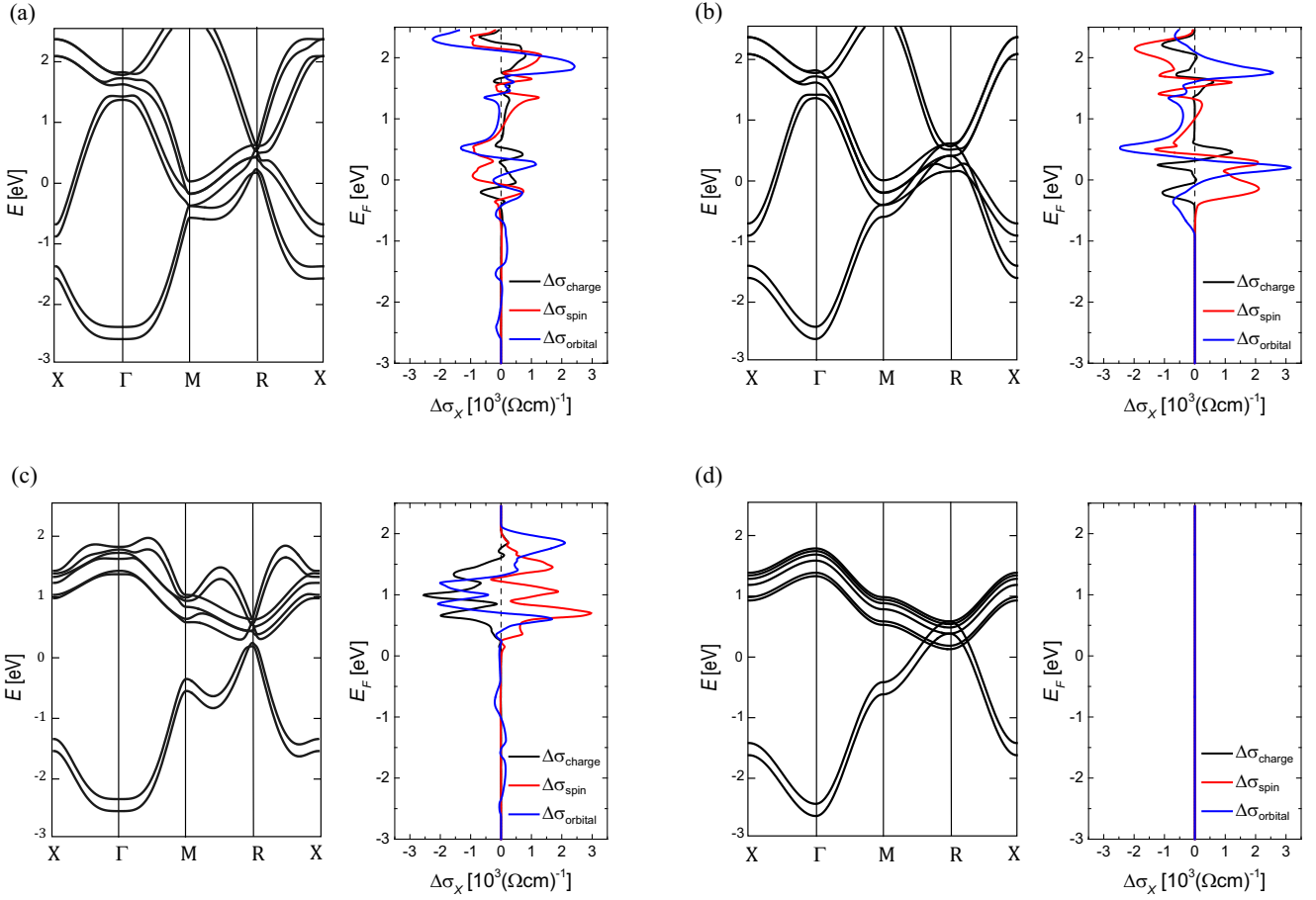}
\caption{
Band structures and corresponding $E_F$ dependence of AMR and OAMR (a) with both orbital anisotropy and hybridization, (b) with orbital anisotropy and without orbital hybridization, (c) without orbital anisotropy and with orbital hybridization, and (d) without both orbital anisotropy and orbital hybridization. (Left panels) The band structures of tight-binding model [Eq.~(\ref{AEq2})] for mangetization along the $z$ direction where the lowest two bands correspond to $s$ orbital state. (Right panels) The AMR of charge, spin, and orbital conductivities in which the black, red, and blue curves correspond to the charge, spin, and orbital conductivities. Note that $\hbar/e$ is defined as unity for convenience.
}
\label{fig:FIG2}
\end{center}
\end{figure*}

In this section, we numerically compute the longitudinal charge, spin, and orbital conductivities based on Green's function formalism~\cite{Ghosh2018}, not relying on the perturbation theory for SOC. Detailed description of the Hamiltonian and its numerical parameters are given in Appendix~\ref{appendix: TB Ham}. Note that the orbital currents carry an angular momentum aligned along the magnetization. We assume an electric field $\mathcal{E}_x$ along the $x$ direction and calculate the resulting conductivities which are composed of the Fermi surface and Fermi sea contributions:

\begin{align}
\sigma_X &= \sigma_X^\RN{1}+\sigma_X^\RN{2}, \nonumber \\
\sigma_X^\RN{1} = &-\frac{\hbar\mathcal{E}_x}{4\pi}\int \frac{d^3\vec{k}}{(2\pi)^3} \nonumber \\
&\times\Tr\left\{Xg^R_Ev_x (g^R_E-g^A_E)-(g^R_E-g^A_E)v_xg^A_E]\right\}_{E_F}, \nonumber \\
\sigma_X^\RN{2} = &-\frac{\hbar\mathcal{E}_x}{4\pi}\int_{-\infty}^{E_F} dE \int \frac{d^3\vec{k}}{(2\pi)^3} \nonumber \\
&\times\Tr\left\{X\left[\frac{\partial{g^R_E}}{\partial{E}}v_xg^R_E-g^R_Ev_x\frac{\partial{g^R_E}}{\partial{E}}-\langle R \leftrightarrow A \rangle\right]\right\}, \label{Eq6}
\end{align}
where the subscript $X$ refers to charge, spin, and orbital, and the corresponding operators $X$ for the physical quantities are $v_x$, $\sigma_\vec{m}\otimes v_x$, and $(L_\vec{m}v_x+v_xL_\vec{m})/2$, respectively. Here $v_x$ is velocity matrix defined as $\partial_{k_x}\cal{H}$ and $\sigma_\vec{m}(L_\vec{m})$ is Pauli matrix (orbital angular momentum operator) projected to the magnetization direction $\vec{m}$. $\sigma_\alpha^\RN{1}$ is the Fermi surface contribution and $\sigma_\alpha^\RN{2}$ is the Fermi sea contribution, $g^{R(A)}_E$ is retarded (advanced) Green function defined as $(E-{\cal{H}}\pm i\eta)^{-1}$ where $\eta$ is the level broadening. We only consider the Fermi surface contribution since the Fermi sea contribution is found to be negligible in our calculation.

We distinguish two orbital factors and their consequential anisotropic conductions. As mentioned in Sec.~\ref{sec: OS}, the orbital anisotropy and the orbital hybridization are controlled via the difference between $p$-to-$p$ hopping parameters $\abs{t_{p\sigma}-t_{p\pi}}$ and the hopping parameter between $s$ and $p$ orbitals $\gamma_{sp}$, respectively. The band structures and the Fermi energy dependence of anisotropic conductivities for each combination of orbital factors are shown in Fig.~\ref{fig:FIG2}. Here, we represent an anisotropic conductance as $\Delta\sigma_X\equiv\sigma_X^{\vec{m}\parallel\vhat{y}}-\sigma_X^{\vec{m}\parallel\vhat{x}}$. 

The transport of electron charge, spin, and orbital shows isotropic behavior on the magnetization direction when both orbital anisotropy and hybridization are absent [Fig.~\ref{fig:FIG2}(d)]. This is due to the exact degeneracy of $p$ orbitals, which yields identical band structures for varied $\vec{m}$. The inclusion of orbital anisotropy or orbital hybridization to the system [Figs.~\ref{fig:FIG2}(a)--\ref{fig:FIG2}(c)] produces AMR indicating a relation between AMR and orbital physics. Besides well-known AMR, which regards charge currents entailing spin degrees of freedom~\cite{Taniguchi2015}, finite $\Delta\sigma_\mathrm{orbital}$ states that the orbital conductivities also depend on the relative orientation between current and magnetization, which we coin as OAMR. This generic and inherent correlation between orbital and magnetization will invigorate the potential utilization of orbital transport~\cite{Go2019,Zhang2019} in ferromagnets. Note that the $s$ orbital may produce  AMR under the presence of orbital hybridization [$-2.5~\mathrm{eV} \le E_F \le 1.0~\mathrm{eV}$ in Figs.~\ref{fig:FIG2}(a) and \ref{fig:FIG2}(c)], through which the anisotropic property of $p$ orbitals is imparted. In addition, there are somewhat complicated resemblance between the $E_F$ dependence of AMR and OAMR. This tendency is comprehensible considering the fact that the SOC correlates the feature of orbital and spin~\cite{Tanaka2008,Kontani2009,Go2018}, where a further examination is done below.

\begin{figure}[t]
\begin{center}
\includegraphics[scale=0.475]{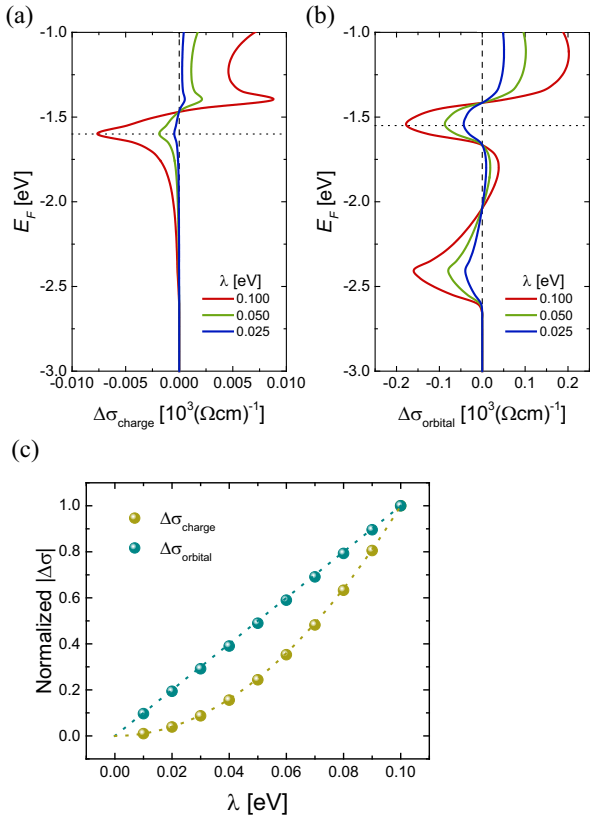}
\caption{
$E_F$ dependence of (a) AMR and (b) OAMR with varied SOC strength, $\lambda$. Horizontal dotted lines indicate the local maxima ($-1.60$ eV and $-1.55$ eV, respectively) and the dependence of charge and orbital conductivities are calculated at this Fermi energy, which is summarized in (c). OAMR and AMR of varied SOC strength is normalized by OAMR and AMR of $\lambda=0.1~\mathrm{eV}$, which are shown as filled symbols. Dotted lines show linear and quadratic guidelines which well describe the tendency of calculated values.
}
\label{fig:FIG3}
\end{center}
\end{figure}

The relevance of OAMR to AMR is observed through their dependence on the strength of SOC (Fig.~\ref{fig:FIG3}) in the system with both orbital hybridization and anisotropy [Fig.~\ref{fig:FIG2}(a)]. To avoid severe alteration of the band structures, we restrict our interest to \textit{s} band contribution and weak SOC regime which gives the spin and orbital as a good quantum number. The calculated AMR and OAMR with $\lambda= \ $0.1, 0.05, 0.025 eV are shown in Fig.~\ref{fig:FIG3}(a) and (b). Note that both AMR and OAMR vanish when $\lambda=0$. Then, we select a specific Fermi energy for each phenomena to precisely determine the order of SOC. For clarity, $-1.60$ eV and $-1.55$ eV are selected repsectively in which AMR and OAMR exhibit local maxima and the corresponding values are normalized by $\Delta\sigma$ at $\lambda = \ $ 0.1 eV. The dotted guidelines show that AMR and OAMR follow quadratic and linear manner with respect to SOC. The linear dependence of OAMR may be understood as a consequence of effective SOC, $\lambda\vec{L}\cdot\vec{m}$, which provides a pathway to interact with the magnetization in terms of magnetization-dependent orbital energies. As long as the anisotropic band structures prevail the electric conduction, the magnetization-dependent orbital currents will provoke spin currents in similar manner according to the SOC. On that ground, the OAMR is appreciated as the new route to AMR. Note that although the strength of SOC is chosen to be comparable to or weaker than that of exchange coupling to maintain consistency with the previous analytical calculation, strong spin-orbit coupling does not alter our main conclusion qualitatively (not shown).

\section{\label{sec: Con}Conclusion}

To conclude, we investigated the orbital origin of AMR based on a tight-binding model with \textit{s} and \textit{p} orbitals. We found that the orbital hybridization is the unnoticed origin of AMR which highlights the role of orbital physics in understanding spin-related phenomena in spin-orbit-coupled systems. Along with the orbital anisotropy, the orbital hybridization generates the momentum-dependent orbital splitting which is the main ingredient of magnetization-dependent band structures. Through the effective correlation between the orbital and the magnetization, the electronic structure of system responds to the magnetization direction, which ultimately derives AMR. We showed that not only charge (or spin) conductivity but also orbital conductivity exhibits magnetization-dependent tendency which we called OAMR. Moreover, this OAMR contributes to AMR through SOC as the orbital Hall effect does for the spin Hall effect~\cite{Tanaka2008,Kontani2009,Go2018}. The alikeness of OAMR and orbital Hall effect is also found in the fact that the orbital hybridization is the common source of both phenomena, emphasizing the importance of the orbital hybridization in several substantial observations of spintronics. In this respect, the measurement of AMR in materials exhibiting considerable orbital Hall conductivities~\cite{Jo2018} is intriguing. For systematic investigation, we also propose non-magnetic metals where AMR and OAMR occur through the Zeeman coupling but material-dependent exchange coupling is excluded.
 
\begin{acknowledgments}
K.-J.L. acknowledges support from the KIST Institutional Program (2V05750), the National Research Foundation of Korea (NRF) funded by the Ministry of Science and ICT (2020R1A2C3013302) and the KU-KIST Graduate School Program. K.-W.K. acknowledges support from the KIST Institutional Program (2E30600), the National Research Council of Science \& Technology (NST) (CAP-16-01-KIST), and the National Research and Development Program through the NRF (2019M3F3A1A02071509). G.G. acknowledges support from NRF (2019R1I1A1A01063594).
\end{acknowledgments}

\appendix
\section{\label{appendix: TB Ham}Tight-binding Hamiltonian}

We consider a general ferromagnetic system with SOC in a simple cubic structure. Since the conduction electrons have $s$ character in most materials under our consideration, the minimal model should include $s$ and $p$ orbitals to realize a mixture between different orbital states. By assuming $s$, $p_x$, $p_y$, and $p_z$ orbitals at each Bravais lattice, the Bloch basis $\ket{\psi_{\alpha\beta}^\vec{k}(\vec{r})}$ are defined by,
\begin{align}
\ket{\psi_{\alpha\beta}^\vec{k}(\vec{r})} = \sum_{\vec{R}} e^{i\vec{k}\cdot\vec{R}}\ket{\psi_{\alpha\beta}(\vec{r})}
\end{align}
where $\vec{R}$ is the lattice sites, $\alpha=s,\ p_x,\ p_y,\ p_z$, and $\beta=\uparrow,\downarrow$. Based on the Bloch basis, the tight-binding Hamiltonian is written as,
\begin{align}
{\cal{H}}={\cal{H}}_\mathrm{L}+\frac{2\Delta_\mathrm{ex}}{\hbar}\vec{m}\cdot\vec{S}+\frac{2\lambda}{\hbar^2}\vec{L}\cdot\vec{S}. \label{AEq2}
\end{align}
Here, $\cal{H}_\mathrm{L}$ is the orbital part of Hamiltonian, $\Delta_\mathrm{ex}$ is the exchange parameter, $\lambda$ is the strength of SOC, and $\vec{m}$, $\vec{S}$, $\vec{L}$ correspond to the unit vector along the magnetization direction, spin angular momentum operator $(\hbar/2)\vec{\sigma}$, and orbital angular momentum operator, respectively. The components of orbital angular momentum operator $\vec{L}$ in the $(\ket{\psi_{s}^\vec{k}(\vec{r})},\ket{\psi_{p_x}^\vec{k}(\vec{r})},\ket{\psi_{p_y}^\vec{k}(\vec{r})},\ket{\psi_{p_z}^\vec{k}(\vec{r})})$ basis is
\begin{align}
L_x=\begin{pmatrix}
0 & 0 & 0 & 0 \\
0 & 0 & 0 & 0 \\
0 & 0 & 0 & -i \\
0 & 0 & i & 0 \\
\end{pmatrix}, \nonumber \\
L_y=\begin{pmatrix}
0 & 0 & 0 & 0 \\
0 & 0 & 0 & i \\
0 & 0 & 0 & 0 \\
0 & -i & 0 & 0 \\
\end{pmatrix}, \nonumber \\
L_z=\begin{pmatrix}
0 & 0 & 0 & 0 \\
0 & 0 & -i & 0 \\
0 & i & 0 & 0 \\
0 & 0 & 0 & 0 \\
\end{pmatrix}.
\end{align}
The second and third terms in Eq.~(\ref{AEq2}) are the exchange interaction and SOC, respectively, and the orbital part of Hamiltonian [Eq.~(\ref{Eq1})] is composed of matrix elements given as,
\begin{align}
\braket{\psi_{s\uparrow(\downarrow)}^\vec{k}|{\cal{H}}_\mathrm{L}|\psi_{s\uparrow(\downarrow)}^\vec{k}} =& \ E_s^\mathrm{on}+2t_{s\sigma}(\cos{k_xa} \nonumber \\
&+\cos{k_ya}+\cos{k_za}), \nonumber \\
\braket{\psi_{p_x\uparrow(\downarrow)}^\vec{k}|{\cal{H}}_\mathrm{L}|\psi_{p_x\uparrow(\downarrow)}^\vec{k}} =& \ E_p^\mathrm{on}+2t_{p\sigma}\cos{k_xa} \nonumber \\
&+2t_{p\pi}(\cos{k_ya}+\cos{k_za}), \nonumber \\
\braket{\psi_{p_y\uparrow(\downarrow)}^\vec{k}|{\cal{H}}_\mathrm{L}|\psi_{p_y\uparrow(\downarrow)}^\vec{k}} =& \ E_p^\mathrm{on}+2t_{p\sigma}\cos{k_ya} \nonumber \\
&+2t_{p\pi}(\cos{k_za}+\cos{k_xa}), \nonumber \\
\braket{\psi_{p_z\uparrow(\downarrow)}^\vec{k}|{\cal{H}}_\mathrm{L}|\psi_{p_z\uparrow(\downarrow)}^\vec{k}} =& \ E_p^\mathrm{on}+2t_{p\sigma}\cos{k_za} \nonumber \\
&+2t_{p\pi}(\cos{k_xa}+\cos{k_ya}), \nonumber \\
\braket{\psi_{s\uparrow(\downarrow)}^\vec{k}|{\cal{H}}_\mathrm{L}|\psi_{p_x\uparrow(\downarrow)}^\vec{k}} =& \ 2i\gamma_{sp}\sin{k_xa}, \nonumber \\
\braket{\psi_{s\uparrow(\downarrow)}^\vec{k}|{\cal{H}}_\mathrm{L}|\psi_{p_y\uparrow(\downarrow)}^\vec{k}} =& \ 2i\gamma_{sp}\sin{k_ya}, \nonumber \\
\braket{\psi_{s\uparrow(\downarrow)}^\vec{k}|{\cal{H}}_\mathrm{L}|\psi_{p_z\uparrow(\downarrow)}^\vec{k}} =& \ 2i\gamma_{sp}\sin{k_za},
\end{align}
where $E_{s(p)}^\mathrm{on}$ is the on-site energy of $s(p)$ orbital, and $t_{s\sigma}$, $t_{p\sigma(\pi)}$, $\gamma_{sp}$ are the two-center integrals~\cite{Slater1954} corresponding to the nearest-neighbor hopping between $s$ orbitals, $p$ orbitals through $\sigma(\pi)$ bond, and between $s$ and $p$ orbitals respectively. Numerical values of parameters are chosen as $E_s^\mathrm{on}=-1.0,\ E_{p}^\mathrm{on}=1.0,\ t_{s\sigma}=-0.25,\ t_{p\sigma}=0.6 \ (0.1),\ t_{p\pi}=-0.15 \ (0.1),\ \gamma_{sp}=0.5,\ \lambda=0.1,\ \Delta_{\rm ex}=0.1$, all in units of eV. Values in parenthesis correspond to the system without orbital anisotropy. Note that $E_s<E_p$ resembles the orbital level of ferromagnetic metals ($3d<4s<4p$).

\section{\label{appendix: Spher Coord}Spherical coordinates}

For a given Bloch momentum $\vec{k}$, one can uniquely define the two angles $\theta_\vec{k}=\cos^{-1}{k_z/\abs{\vec{k}}}$ and $\phi_\vec{k}=\arg{(k_x+ik_y)}$. Then, we can define the three unit vectors $(\hat{\vec{k}},\vhat{\theta}_\vec{k},\vhat{\phi}_\vec{k})$ for a spherical coordinate system, where
\begin{align}
\hat{\vec{k}}&=(\sin{\theta_\vec{k}}\cos{\phi_\vec{k}},\sin{\theta_\vec{k}}\sin{\phi_\vec{k}},\cos{\theta_\vec{k}}),\nonumber\\
\vhat{\theta}_\vec{k}&=(\cos{\theta_\vec{k}}\cos{\phi_\vec{k}},\cos{\theta_\vec{k}}\sin{\phi_\vec{k}},-\sin{\theta_\vec{k}}),\nonumber\\
\vhat{\phi}_\vec{k}&=(-\sin{\phi_\vec{k}},\cos{\phi_\vec{k}},0).\label{BEq1}
\end{align}
Let $\ket{\vec{kp}^\sigma}=(\ket{\vec{k}p_x^\sigma},\ket{\vec{k}p_y^\sigma},\ket{\vec{k}p_z^\sigma})$ be a three dimensional vector consisting of $p$ orbital states where $\sigma=\ \uparrow,\downarrow$. Note that $\ket{\vec{k}\alpha^\beta}\equiv\ket{\psi_{\alpha\beta}^\vec{k}}$ where $\alpha=s,\ p_x,\ p_y,\ p_z$ and $\beta=\ \uparrow,\downarrow$. Then $\ket{\vec{k}p_\vec{k}^\sigma}=\ket{\vec{kp}^\sigma}\cdot\hat{\vec{k}}$ is the radial orbital and $\ket{\vec{k}p_{\theta_\vec{k}}^\sigma}=\ket{\vec{kp}^\sigma}\cdot\vhat{\theta}_\vec{k}$ and $\ket{\vec{k}p_{\phi_\vec{k}}^\sigma}=\ket{\vec{kp^\sigma}}\cdot\vhat{\phi}_\vec{k}$ are tangential orbitals. Therefore the orbital part of Hamiltonian is written as,
\begin{align}
{\cal{H}}_\mathrm{L}= &E_s(\vec{k})\ket{\vec{k}s^\sigma}\bra{\vec{k}s^\sigma}+E_{pr}(\vec{k})\ket{\vec{k}p_\vec{k}^\sigma}\bra{\vec{k}p_\vec{k}^\sigma} \nonumber \\
&+E_{pt}(\vec{k})(\ket{\vec{k}p_{\theta_\vec{k}}^\sigma}\bra{\vec{k}p_{\theta_\vec{k}}^\sigma}+\ket{\vec{k}p_{\phi_\vec{k}}^\sigma}\bra{\vec{k}p_{\phi_\vec{k}}^\sigma}).
\end{align}

\begin{comment}
For $k_z=0$, the description of radial and tangential $p$-orbital states is simplified as,
\begin{align}
&\ket{\vec{k}p_\vec{k}^\sigma}_{k_z=0} = \cos{\phi_\vec{k}}\ket{\vec{k}p_x^\sigma}+\sin{\phi_\vec{k}}\ket{\vec{k}p_y^\sigma}, \nonumber \\
&\ket{\vec{k}p_{\theta_\vec{k}}^\sigma}_{k_z=0} = -\ket{\vec{k}p_z^\sigma}, \nonumber \\
&\ket{\vec{k}p_{\phi_\vec{k}}^\sigma}_{k_z=0} = -\sin{\phi_\vec{k}}\ket{\vec{k}p_x^\sigma}+\cos{\phi_\vec{k}}\ket{\vec{k}p_y^\sigma}. \label{BEq2}
\end{align}
\end{comment}

The orbital angular momentum operator $\vec{L}$ is also rewritten in spherical coordinates as,
\begin{align}
&\vec{L}\cdot\hat{\vec{k}}=i\hbar\ket{\vec{k}p_{\phi_\vec{k}}^\sigma}\bra{\vec{k}p_{\theta_\vec{k}}^\sigma}-i\hbar\ket{\vec{k}p_{\theta_\vec{k}}^\sigma}\bra{\vec{k}p_{\phi_\vec{k}}^\sigma}, \nonumber \\
&\vec{L}\cdot\vhat{\theta}_\vec{k}=i\hbar\ket{\vec{k}p_\vec{k}^\sigma}\bra{\vec{k}p_{\phi_\vec{k}}^\sigma}-i\hbar\ket{\vec{k}p_{\phi_\vec{k}}^\sigma}\bra{\vec{k}p_\vec{k}^\sigma}, \nonumber \\
&\vec{L}\cdot\vhat{\phi}_\vec{k}=i\hbar\ket{\vec{k}p_{\theta_\vec{k}}^\sigma}\bra{\vec{k}p_\vec{k}^\sigma}-i\hbar\ket{\vec{k}p_\vec{k}^\sigma}\bra{\vec{k}p_{\theta_\vec{k}}^\sigma}, \label{BEq3}
\end{align}
which gives the effective SOC in the basis $(\ket{\vec{k}s^\sigma},\ket{\vec{k}p_\vec{k}^\sigma},\ket{\vec{k}p_{\theta_\vec{k}}^\sigma},\ket{\vec{k}p_{\phi_\vec{k}}^\sigma})$ as,
\begin{align}
\cal{H}_\mathrm{SO}&= -\frac{\lambda}{\hbar}\vec{L}\cdot\vec{m} \nonumber \\
&=-\lambda\begin{pmatrix}
0 & 0 & 0 & 0 \\
0 & 0 & -im_{\phi_\vec{k}} & im_{\theta_\vec{k}} \\
0 & im_{\phi_\vec{k}} & 0 & -im_\vec{k} \\
0 & -im_{\theta_\vec{k}} & im_\vec{k} & 0 \\
\end{pmatrix}, \label{BEq4}
\end{align}
where $\vec{m}=m_\vec{k}\hat{\vec{k}}+m_{\theta_\vec{k}}\vhat{\theta}_\vec{k}+m_{\phi_\vec{k}}\vhat{\phi}_\vec{k}$.

Another useful basis is given by a linear combination of the tangential orbitals. Defining $\ket{\vec{k}p_\pm^\sigma}=(\ket{\vec{k}p_{\theta_\vec{k}}^\sigma}\pm i\ket{\vec{k}p_{\phi_\vec{k}}^\sigma})/\sqrt{2}$, the off-diagonal components in the tangential orbitals are eliminated. In the basis $(\ket{\vec{k}s^\sigma},\ket{\vec{k}p_\vec{k}^\sigma},\ket{\vec{k}p_-^\sigma},\ket{\vec{k}p_+^\sigma})$,
\begin{equation}
\cal{H}_\mathrm{SO}=-\lambda\begin{pmatrix}
0 & 0 & 0 & 0 \\
0 & 0 & m_{-,\vec{k}} & im_{+,\vec{k}} \\
0 & m_{+,\vec{k}} & -m_\vec{k} & 0 \\
0 & -im_{-,\vec{k}} & 0 & m_\vec{k} \\
\end{pmatrix}, \label{BEq5}
\end{equation}
%\vspace{1.3px}
where $m_{\pm,\vec{k}}=(m_{\theta_\vec{k}}\pm im_{\phi_\vec{k}})/\sqrt{2}$.

\begin{widetext}
\section{\label{appendix: exact S}Exact expression of $\mathcal{S}$}

The exact expression of $\mathcal{S}$ satisfying $[\mathcal{H}_0,\mathcal{S}]=\mathcal{H}_1$ without small $\lambda$ approximation is given by
\begin{align}
\mathcal{S}&=\begin{pmatrix}
0 & S_{2\times2} \\ 
-S_{2\times2}^\dagger & 0
\end{pmatrix} ,\nonumber\\
S_{2\times2}&=\lambda \begin{pmatrix}
\displaystyle\frac{i\gamma_{sp}m_{-,\vec{k}}}{(E_{pr}-E_{pt}-\lambda m_\vec{k})(E_{pt}-E_s+\lambda m_\vec{k})+\gamma_{sp}^2}&\displaystyle -\frac{\gamma_{sp}m_{+,\vec{k}}}{(E_{pr}-E_{pt}+\lambda m_\vec{k})(E_{pt}-E_s-\lambda m_\vec{k})+\gamma_{sp}^2} \\ 
-\displaystyle\frac{m_{-,\vec{k}}(E_{pt}-E_s+\lambda m_\vec{k})}{(E_{pr}-E_{pt}-\lambda m_\vec{k})(E_{pt}-E_s+\lambda m_\vec{k})+\gamma_{sp}^2}&\displaystyle -\frac{im_{+,\vec{k}}(E_{pt}-E_s-\lambda m_\vec{k})}{(E_{pr}-E_{pt}+\lambda m_\vec{k})(E_{pt}-E_s-\lambda m_\vec{k})+\gamma_{sp}^2}
\end{pmatrix}.
\end{align}
\end{widetext}

\section{\label{appendix: AMR}Derivation of Eq.~(\ref{Eq:AMR})}
We rewrite Eq.~(\ref{Eq:sigma}) as
\begin{equation}
\sigma=\frac{e^2\tau}{V}\Im\sum_{\vec{k}}\frac{v_{0,x}(\vec{k})^2}{E_F-E_0(\vec{k})-i\epsilon},
\end{equation}
for an infinitesimal $\epsilon>0$. Here we take the single $s$ band ($n=0$) only as mentioned in the main text. We assume that $E_{0,0}(\vec{k})$ and $E_{0,2}(\vec{k})$ are spherically symmetric and quadratic in $k$. Then, we may write
\begin{align}
E_0(\vec{k})&=ak^2+b(\vec{m}\cdot\vec{k})^2,\\
v_{0,x}(\vec{k})&=\frac{2a}{\hbar}k_x+\frac{2b}{\hbar} m_x(\vec{m}\cdot\vec{k}),
\end{align}
where $E_{0,0}(k)=ak^2$ and $E_{0,2}(k)=bk^2$ for a positive $a$. Since $b\propto\lambda^2\gamma_{sp}^2$, we keep first order contributions from $b$ only. After some algebra,
\begin{align}
\sigma&=\frac{e^2\tau}{V}\Im\sum_\vec{k}\frac{4 ak^2}{3\hbar^2}\frac{a+2b m_x^2}{E_F-ak^2-i\epsilon}\nonumber\\
&=\frac{4ae^2\tau}{3\hbar^2V}(a+2b m_x^2)\sum_\vec{k}k^2\delta(E_F-ak^2)\nonumber\\
&=\frac{2ae^2\tau}{3\hbar^2\pi^2}(a+2b m_x^2)\int dk k^4\delta(E_F-ak^2)\nonumber\\
&=\frac{e^2\tau}{3\hbar^2\pi^2}(a+2b m_x^2)\int dk k^3\delta(k_F-k)\nonumber\\
&=\frac{e^2\tau}{3\hbar^2\pi^2}(a+2b m_x^2)k_F^3\nonumber\\
&=\frac{e^2\tau k_F}{3\hbar^2\pi^2}[E_F+2E_{0,2}(k_F) m_x^2],
\end{align}
where $k_F=\sqrt{E_F/a}$.

%\bibliography{C:/Users/john/Desktop/APS/APS/ref}

\begin{thebibliography}{99}
%
\bibitem{Thomson1857} W. Thomson, Proc. R. Soc. Lond. \textbf{8}, 546 (1857).
%
\bibitem{Smit1951} J. Smit, Physica \textbf{17}, 612 (1951).
%
\bibitem{Campbell1969} I. A. Campbell, A. Fert, and O. Jaoul, J. Phys. C: Solid State Phys. \textbf{3}, S95 (1970).
%
\bibitem{Berger1964} L. Berger, Physica \textbf{30}, 1141 (1964).
%
\bibitem{Mcguire1975} T. R. McGuire and R. Potter, IEEE Trans. Magn. \textbf{11}, 1018 (1975).
%
\bibitem{Velev2005} J. Velev, R. F. Sabirianov, S. S. Jaswal, and E. Y. Tsymbal, Phys. Rev. Lett. \textbf{94}, 127203 (2005).
%
\bibitem{Hsieh2009a} D. Hsieh, Y. Xia, L. Wray, D. Qian, A. Pal, J. H. Dil, J. Osterwalder, F. Meier, G. Bihlmayer, C. L. Kane, Y. S. Hor, R. J. Cava, and M. Z. Hasan, Science \textbf{323}, 919 (2009)
%
\bibitem{Hsieh2009b} D. Hsieh, Y. Xia, D. Qian, L. Wray, J. H. Dil, F. Meier, J. Osterwalder, L. Patthey, J. G. Checkelsky, N. P. Ong, A. V. Fedorov, H. Lin, A. Bansil, D. Grauer, Y. S. Hor, R. J. Cava, and M. Z. Hasan, Nature \textbf{460}, 1101 (2009).
%
\bibitem{Edelstein1990} V. M. Edelstein, Solid State Commun. \textbf{73}, 233 (1990).
%
\bibitem{Roushan2009} P. Roushan, J. Seo, C. V. Parker, Y. S. Hor, D. Hsieh, D. Qian, A. Richardella, M. Z. Hasan, R. J. Cava, and A. Yazdani, Nature \textbf{460}, 1106 (2009).
%
\bibitem{Park2012} S. R. Park, J. Han, C. Kim, Y. Y. Koh, C. Kim, H. Lee, H. J. Choi, J. H. Han, K. D. Lee, N. J. Hur, M. Arita, K. Shimada, H. Namatame, and M. Taniguchi, Phys. Rev. Lett. \textbf{108}, 046805 (2012).
%
\bibitem{Xie2014} Z. Xie, S. He, C. Chen, Y. Feng, H. Yi, A. Liang, L. Zhao, D. Mou, J. He, Y. Peng, X. Liu, Y. Liu, G. Liu, X. Dong, L. Yu, J. Zhang, S. Zhang, Z. Wang, F. Zhang, F. Yang, Q. Peng, X. Wang, C. Chen, Z. Xu, and X. J. Zhou, Nat. Commun. \textbf{5}, 3382 (2014).
%
\bibitem{Zhang2013} H. Zhang, C.-X. Liu, and S.-C. Zhang, Phys. Rev. Lett. \textbf{111}, 066801 (2013).
%
\bibitem{Park2011} S. R. Park, C. H. Kim, J. Yu, J. H. Han, and C. Kim, Phys. Rev. Lett. \textbf{107}, 156803 (2011).
%
\bibitem{Go2017} D. Go, J.-P. Hanke, P. M. Buhl, F. Freimuth, G. Bihlmayer, H.-W. Lee, Y. Mokrousov, and S. Bl\"{u}gel, Sci. Rep. \textbf{7}, 46742 (2017).
%
\bibitem{Tanaka2008} T. Tanaka, H. Kontani, M. Naito, T. Naito, D. S. Hirashima, K. Yamada, and J. Inoue, Phys. Rev. B \textbf{77}, 165117 (2008).
%
\bibitem{Kontani2009} H. Kontani, T. Tanaka, D. S. Hirashima, K. Yamada, and J. Inoue, Phys. Rev. Lett. \textbf{102}, 016601 (2009).
%
\bibitem{Go2018} D. Go, D. Jo, C. Kim, and H.-W. Lee, Phys. Rev. Lett. \textbf{121}, 086602 (2018).
%
\bibitem{Paul} P. M. Haney and M. D. Stiles, Phys. Rev. Lett. \textbf{105}, 126602 (2010).
%
\bibitem{Ghosh2018} S. Ghosh and A. Manchon, Phys. Rev. B \textbf{97}, 134402 (2018).

\bibitem{SWolff}  J. R. Schrieffer and P. A. Wolff, Phys. Rev. \textbf{149}, 491 (1966).
%
\bibitem{Taniguchi2015} T. Taniguchi, J. Grollier, and M. D. Stiles, Phys. Rev. Applied \textbf{3}, 044001 (2015).
%
\bibitem{Go2019} D. Go and H.-W. Lee, Phys. Rev. Research \textbf{2}, 013177 (2020).
%
\bibitem{Zhang2019} L. Zhang, F. R. Lux, J.-P. Hanke, P. M. Buhl, S. Grytsiuk, S. Bl\"{u}gel, and Y. Mokrousov, arXiv preprint arXiv: 1910.03317 (2019).
%
\bibitem{Slater1954} J. C. Slater and G. F. Koster, Phys. Rev. \textbf{94}, 1498 (1954).
%
\bibitem{Jo2018} D. Jo, D. Go, and H.-W. Lee, Phys. Rev. B \textbf{98}, 214405 (2018).
%
%\bibitem{Cao2013} Y. Cao, J. A. Waugh, X-W. Zhang, J-W. Luo, Q. Wang, T. J. Reber, S. K. Mo, Z. Xu, A. Yang, J. Schneeloch, G. D. Gu, M. Brahlek, N. Bansal, S. Oh, A. Zunger, and D. S. Dessau, Nat. Phys. \textbf{9}, 499 (2013).
%
%\bibitem{Zeljkovic2014} I. Zeljkovic, Y. Okada, C.-Y. Huang, R. Sankar, D. Walkup, W. Zhou, M. Serbyn, F. Chou, W.-F. Tsai, H. Lin, A. Bansil, L. Fu, M. Zahid Hasan, and V. Madhavan, Nat. Phys. \textbf{10}, 572 (2014).
\end{thebibliography}
\end{document}